\documentclass[reprint,aps,prl,amsmath,amssymb,superscriptaddress, preprintnumbers]{revtex4-2}
 
\usepackage[utf8]{inputenc}
\usepackage[T1]{fontenc}

\usepackage{amsmath}
\usepackage{mathrsfs}
\usepackage{txfonts}
\usepackage{mathtools}
\usepackage{braket}
\usepackage{tensor}
\usepackage{xcolor}
\usepackage[abbreviations]{siunitx}
\usepackage{graphicx}
\usepackage{booktabs}

\usepackage{float}

\usepackage{braket}
\usepackage{ulem}

\usepackage{hyperref}
\usepackage{cleveref}

\newcommand{\beq}{\begin{equation}}
\newcommand{\eeq}{\end{equation}}
\newcommand{\beqn}{\begin{eqnarray}}
\newcommand{\eeqn}{\end{eqnarray}}
\newcommand{\bsub}{\begin{subequations}}
\newcommand{\esub}{\end{subequations}}
\newcommand{\bpm}{\begin{pmatrix}}
\newcommand{\epm}{\end{pmatrix}}

\begin{document}
     
\title{From spin to pseudospin symmetry: The origin of magic numbers in nuclear structure}
   
\author{C.~R.~Ding}  
\affiliation{School of Physics and Astronomy, Sun Yat-sen University, Zhuhai 519082, P.R. China.}
\affiliation{Guangdong Provincial Key Laboratory of Quantum Metrology and Sensing, Sun Yat-Sen University, Zhuhai 519082, China }

\author{C.~C.~Wang}
\affiliation{School of Physics and Astronomy, Sun Yat-sen University, Zhuhai 519082, P.R. China.}
\affiliation{Graduate School of China Academy of Engineering Physics, Beijing 100193, China}

\author{J.~M.~Yao}
\email{Corresponding author: yaojm8@sysu.edu.cn}
\affiliation{School of Physics and Astronomy, Sun Yat-sen University, Zhuhai 519082, P.R. China.}
\affiliation{Guangdong Provincial Key Laboratory of Quantum Metrology and Sensing, Sun Yat-Sen University, Zhuhai 519082, China } 

\author{H.~Hergert}
\email{Corresponding author: hergert@frib.msu.edu}
\affiliation{Facility for Rare Isotope Beams, Michigan State University, East Lansing, Michigan 48824-1321, USA.}
\affiliation{Department of Physics \& Astronomy, Michigan State University, East Lansing, Michigan 48824-1321, USA.}

\author{H. Z. Liang}
\affiliation{Department of Physics, Graduate School of Science, The University of Tokyo, Tokyo 113-0033, Japan} 

\author{S.~K.~Bogner}
\affiliation{Facility for Rare Isotope Beams, Michigan State University, East Lansing, Michigan 48824-1321, USA.}
\affiliation{Department of Physics \& Astronomy, Michigan State University, East Lansing, Michigan 48824-1321, USA.}

\date{\today}

\begin{abstract}
Magic numbers lie at the heart of nuclear structure, reflecting enhanced stability in nuclei with closed shells. While the emergence of magic numbers beyond 20 is commonly attributed to strong spin-orbit coupling, the microscopic origin of the spin-orbit potential remains elusive, owing to its dependence on the resolution scale and renormalization scheme of nuclear forces. Here, we investigate the evolution of shell structure with varying momentum resolution in nuclear interactions derived from chiral effective field theory, using the similarity renormalization group to link different scales. We uncover a novel transition from spin symmetry to pseudospin symmetry as the resolution scale decreases, during which magic numbers emerge naturally. A similar pattern is found in calculations using relativistic one-boson-exchange potentials, underscoring the robustness of the phenomenon. This establishes a direct connection between realistic nuclear forces with a high resolution scale and effective nuclear forces at coarse-grained scales, offering a first-principles explanation for the origin of magic numbers and pseudospin symmetry in nuclear shell structure, and new insights into the structure of exotic nuclei far from stability.
\end{abstract}

\preprint{}

\maketitle

 
\textit{\textbf{Introduction.}}  
The concepts of shell structure and magic numbers are central to both atomic and nuclear physics, underpinning the stability and behavior of quantum many-body systems. In atomic physics, electron shell filling governs chemical properties~\cite{Stoner1924,Pauli:1925}, while in nuclear physics the shell model was introduced to explain irregular trends in nuclear observables with nucleon number~\cite{Mayer:1948,Feenberg:1949,Nordheim:1949}. Notably, nuclei with neutron or proton numbers $2, 8, 20, 28, 50, 82$, and $126$ exhibit pronounced discontinuities in binding energies and elevated excitation energies~\cite{Ring:1980}. These features were first explained within a mean-field framework by introducing a strong spin–orbit (SO) coupling term into the nucleon single-particle Hamiltonian~\cite{Mayer:1949,Haxel:1949}. As a result, the energy splitting between spin doublets with quantum numbers $(n,\ell,j=\ell+1/2)$ and $(n,\ell,j=\ell-1/2)$ becomes large and strongly breaks spin symmetry in the single-nucleon spectrum. This approach was later extended to describe deformed nuclei~\cite{Nilsson:1955fn,Nilsson:1969zz} by adding a term proportional to the square of the orbital angular momentum, which lowers the energy of states with high $\ell$. Meanwhile, single-particle states with quantum numbers ($n, \ell, j=\ell+1/2$) and ($n-1, \ell+2, j=\ell+3/2$) were found to be nearly degenerate, leading to the introduction of a “pseudo” orbital angular momentum $\tilde\ell=\ell+1$ and the concept of pseudospin symmetry~\cite{Arima:1969,Hecht:1969}. These developments provided a foundational framework for understanding a wide range of low-energy nuclear phenomena~\cite{Bohr:1998v1,Bohr:1998v2}. Despite this well-established picture of nuclear structure, the detailed microscopic origins of the strong SO potential and pseudospin symmetry from the fundamental strong interaction are not fully understood.  
Using quantum chromodynamics (QCD) to directly model nuclei in terms of quarks and gluons is not feasible due to the nonperturbative nature of the strong force at low energies. Nuclear structure theory therefore treats nuclei as quantum many-body systems of point-like nucleons, an approach justified by the effective confinement of quarks and gluons. In nonrelativistic frameworks, shell structures are described using effective nuclear forces or energy density functionals (EDFs), where a strong SO term is introduced phenomenologically to reproduce observed properties~\cite{Negele:1970,Vautherin:1972,Otsuka:2020RMP}. Alternatively, relativistic mean-field (RMF) models~\cite{Miller:1972,Walecka:1974,Brockmann:1978,Walecka:1986} describe nucleons interacting via meson exchange, with nucleon wave functions represented by Dirac spinors. In this framework, the SO potential naturally arises and can be reasonably explained in terms of strong attractive scalar and repulsive vector mean fields~\cite{Reinhard:1989,Ring:1996,Meng:2005PPNP}. Moreover, the appearance of pseudospin symmetry in nucleon single-particle spectra~\cite{Ginocchio:1997,Ginocchio:1998,Meng:1998,Liang:2015,Heitz:2025} and spin symmetry in antiparticle spectra~\cite{Zhou:2003} also find a natural explanation. Despite these successes, the relativistic framework remains phenomenological, as the parameters of the meson-nucleon couplings are fitted directly to observables for finite nuclei rather than nucleon-nucleon scattering data. Consequently, the key open questions remain: How are the phenomenological effective nuclear potentials linked to realistic nuclear forces, and how can we explain the emergence of the observed shell structure in low-energy nuclear systems from first principles? 
  
Chiral effective field theory offers a powerful framework for deriving realistic nuclear forces from first principles~\cite{Weinberg:1979PRL,Weinberg:1991,Epelbaum:2009RMP,Machleidt:2011PR,Lu:2025}. Here, we aim to explore whether by starting from these nuclear forces and employing nuclear \textit{ab initio} methods we can explain the large SO potential in nuclei, uncover the origin of the ``magic numbers'' and pseudospin symmetry, and predict the shell structure of neutron-rich nuclei. To this end, we investigate the evolution of nuclear shell structure as a function of the momentum resolution of nuclear forces, employing similarity renormalization group (SRG)~\cite{Glazek:1993,Wegner:1994,Bogner:2010,Hebeler:2021PR} and in-medium SRG (IMSRG)~\cite{Tsukiyama11,Hergert:2016jk}. The former is used to decouple low- and high-momentum states, while the latter is used to embed particle-hole correlations into the many-body Hamiltonian in a controlled way, connecting the initial and in-medium effective interactions. 
Our results illustrate the transition from spin symmetry to pseudospin symmetry with decreasing momentum resolution, during which nuclear magic numbers emerge naturally. We also find an enhanced impact of three-nucleon forces on the SO splitting as the momentum scale decreases, underlining their importance for a correct description of nuclear shell structure in low-resolution schemes (see, e.g., \cite{Kaiser:2003ux,Kaiser:2008oc,Otsuka:2010prl,Holt:2012,Hagen:2014,Hergert:2013ij,Cipollone:2015fk}). Finally, we demonstrate the emergence of a similar picture in a relativistic Hartree-Fock (RHF) study based on one-boson-exchange potentials (OBEPs) with different momentum cutoffs.


\textit{\textbf{Nuclear chiral interactions with different momentum resolutions.}} We start from an intrinsic nuclear Hamiltonian composed of two-body ($NN$) and three-body ($3N$) interactions,
\begin{equation}
\label{Eq:H}
H_0 = \sum_{i<j} \dfrac{(\mathbf{p}_i-\mathbf{p}_j)^2}{2MA} + \sum_{i<j} V_{ij}^{(NN)} +  \sum_{i<j<k} W_{ijk}^{(3N)},
\end{equation}
where $M$ is the nucleon mass, $A$ the mass number and $\mathbf{p}_i$ the momentum of the $i$-th nucleon. This Hamiltonian is evolved with the SRG \cite{Bogner:2010},
\begin{equation}  
\label{Eq:FlowH}
  \frac{dH_{\lambda}}{d\lambda}=-\frac{4}{\lambda^5}[[T_{\rm rel}, H_\lambda], H_\lambda].
\end{equation}
Here, we use scattering units $\hbar^2/M=1$ and $T_{\rm rel}$ is the relative kinetic energy of two nucleons. Three sets of chiral interactions are used in this work.  For the {\tt NN+3N(lnl)}~\cite{Soma:2020} and the {\tt N2LOGO(394)}~\cite{Jiang:2020} interactions,  the SRG evolution is carried out consistently on both $NN$ and $3N$ parts, with the flow parameter or resolution scale $\lambda$ ranging from infinity to $1.8$ fm$^{-1}$.  For the {\tt EM} family of interactions, the $NN$ part is evolved via the SRG with resolution scales ranging from $\lambda = 2.8$ fm$^{-1}$ to $\lambda = 1.8$ fm$^{-1}$, while the $3N$ low-energy constants (LECs), $c_D$ and $c_E$, are fit to the \nuclide[3]{H} binding energy and the \nuclide[4]{He} matter radius~\cite{Hebeler:2011,Nogga:2004il}.

\textit{\textbf{From spin to pseudospin symmetries and emergence of the magic numbers.}} Nuclear shell structure is characterized by the effective single-particle energies (ESPEs) of nucleons. Experimentally, one can define ESPEs as energy differences between closed-shell nuclei and (low-lying) states of their neighbors, but these are many-body quantities that cannot be attributed to individual nucleons in an unambiguous fashion ~\cite{Cohen:1963RMP,Muthukrishnan:1965,Baranger:1970}. Establishing a theoretical link requires the choice of a many-body scheme, e.g., a mean-field or beyond-mean-field method, whose validity is in turn inextricably linked to the resolution scale and renormalization scheme for the nuclear interactions \cite{Duguet:2012,Duguet:2015}. Still, ESPEs are an important tool for interpreting nuclear structure, especially in an inherently low-resolution picture like the nuclear shell model. 

\begin{figure*}[tb]
    \centering
    \includegraphics[width=1.6\columnwidth]{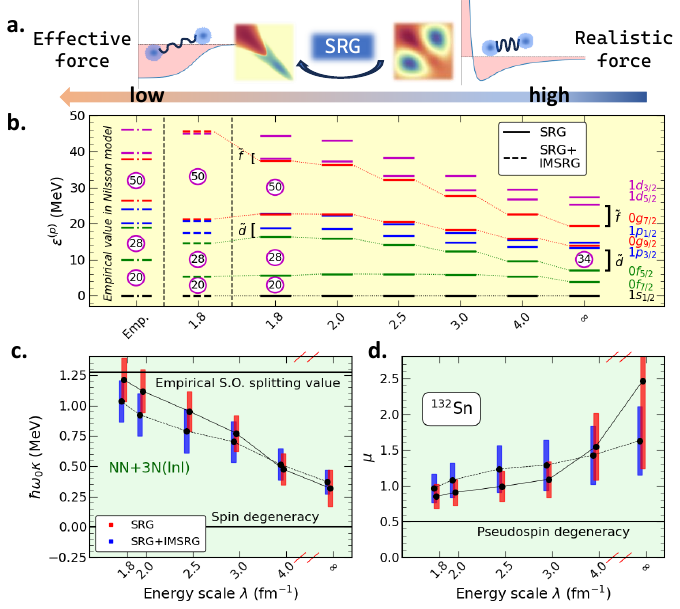}
    \caption{\textbf{Nuclear shell structure in \nuclide[132]{Sn} for the chiral {\tt NN+3N(lnl)} interaction.}  \textbf{a,} Schematic illustration of the nuclear potentials in the ${}^3S_1$ channel in both momentum space and coordinate space, with the momentum resolution scale varying from high to low regimes. \textbf{b,} The ESPEs of protons as functions of $\lambda$, relative to the $1s_{1/2}$ state. The leftmost column shows Nilsson model results with empirical values for comparison. The labels $\tilde{d}$ and $\tilde{f}$ indicate states with $\tilde{\ell}=2$ and $3$, respectively.
  \textbf{c,d,}  Strength parameters $(\hbar\omega_0\kappa, \mu)$ of the Nilsson model, determined by mapping to the results of the chiral {\tt NN+3N(lnl)} interaction at different energy scales. The error bars represent the statistical uncertainty in the fitting process. See the main text for details.
    }
    \label{fig:NN3Nlnl}
\end{figure*}

Here, we perform a series of Hartree-Fock (HF) calculations for doubly magic nuclei using the SRG-evolved $H_\lambda$ with a continuously varying resolution scale $\lambda$ introduced above. The results are labeled as {\tt SRG} in Fig.~\ref{fig:NN3Nlnl}(a), which shows the ESPEs of \nuclide[132]{Sn} as a representative example. To further incorporate nuclear dynamical correlations, we employ the IMSRG approach within the normal-ordered two-body (NO2B) approximation~\cite{Hergert:2016,Hergert:2016jk}. To estimate the effects of improved truncations \cite{Heinz:2021,He:2024}, we performed noniterative checks of the next-order corrections and found them to be small \cite{He:2025}.
The corresponding results are labeled as {\tt SRG+IMSRG}. In both types of calculations, the ESPEs are obtained by diagonalizing the single-particle Hamiltonian, see Eq.~(7) in the Supplemental Material (SM) \footnote{\label{supp}See Supplemental Material at [URL will be inserted by publisher], which includes the additional references \cite{Entem:2003,Bogner:2007,Jurgenson:2009PRL,Hebeler:2012PRC,Miyagi:2023,Wang:2021AME,Wang:2025}.}. 
With SRG alone, the evolution of the initial nuclear force from a high resolution scale to a low-momentum effective interaction significantly increases the SO splittings.  This trend is consistent with findings in previous studies \cite{Pieper:1993,Roth:2006lr,Hagen:2010ll,Cipollone:2015fk,Soma:2015hba}.
As a result, the pronounced shell gaps or magic numbers change from $Z=34$ to $Z=28$ and $Z=50$.

Quantitatively, we can map the ESPEs obtained with chiral interactions at different $\lambda$ onto those of the Nilsson model~\cite{Nilsson:1955fn,Ring:1980} for the spherical case, with $\hbar\omega_0 = 20$ MeV matching our HF and IMSRG calculations. The Nilsson single-particle Hamiltonian consists of the kinetic energy term, a deformed HO potential, as well as spin-orbit and orbital angular momentum terms, $C\mathbf{l}\cdot\mathbf{s} + D \mathbf{l}^2$, where the constants are defined as $C = -2\kappa\hbar\omega_0$ and $D = -\kappa\mu\hbar\omega_0$. The values of the parameters $\kappa$ and $\mu$ obtained for the ESPEs at $\lambda = 1.8$ fm$^{-1}$ are close to their empirical ones, i.e., $\kappa = 0.0637$ and $\mu = 0.6$, for nuclei in the $Z,N\geq 50$ region \cite{Ring:1980}, as shown in Fig.~\ref{fig:NN3Nlnl}(c) and (d) and the SM. Furthermore, Fig.\ref{fig:NN3Nlnl}(b) illustrates that the pseudospin orbital splitting decreases with $\lambda$, indicating the emergence of pseudospin symmetry in the low-resolution nuclear shell structure. Notably, Fig~\ref{fig:NN3Nlnl}(d) demonstrates that the value of $\mu$ approaches $\mu = 0.5$, where the pseudospin doublets become degenerate~\cite{Bahri:1992}. In short, Fig.~\ref{fig:NN3Nlnl} illustrates a transition from spin to pseudospin symmetry as $\lambda$ decreases, causing the emergence of both traditional magic numbers and pseudospin symmetry in the single-nucleon spectrum. We note that genuine observables, such as ground-state energies, remain invariant under changes in the resolution scale (see \cite{Duguet:2015,Soma:2015hba} and SM).

\begin{figure}[bt]
\centering 
\includegraphics[width=\linewidth]{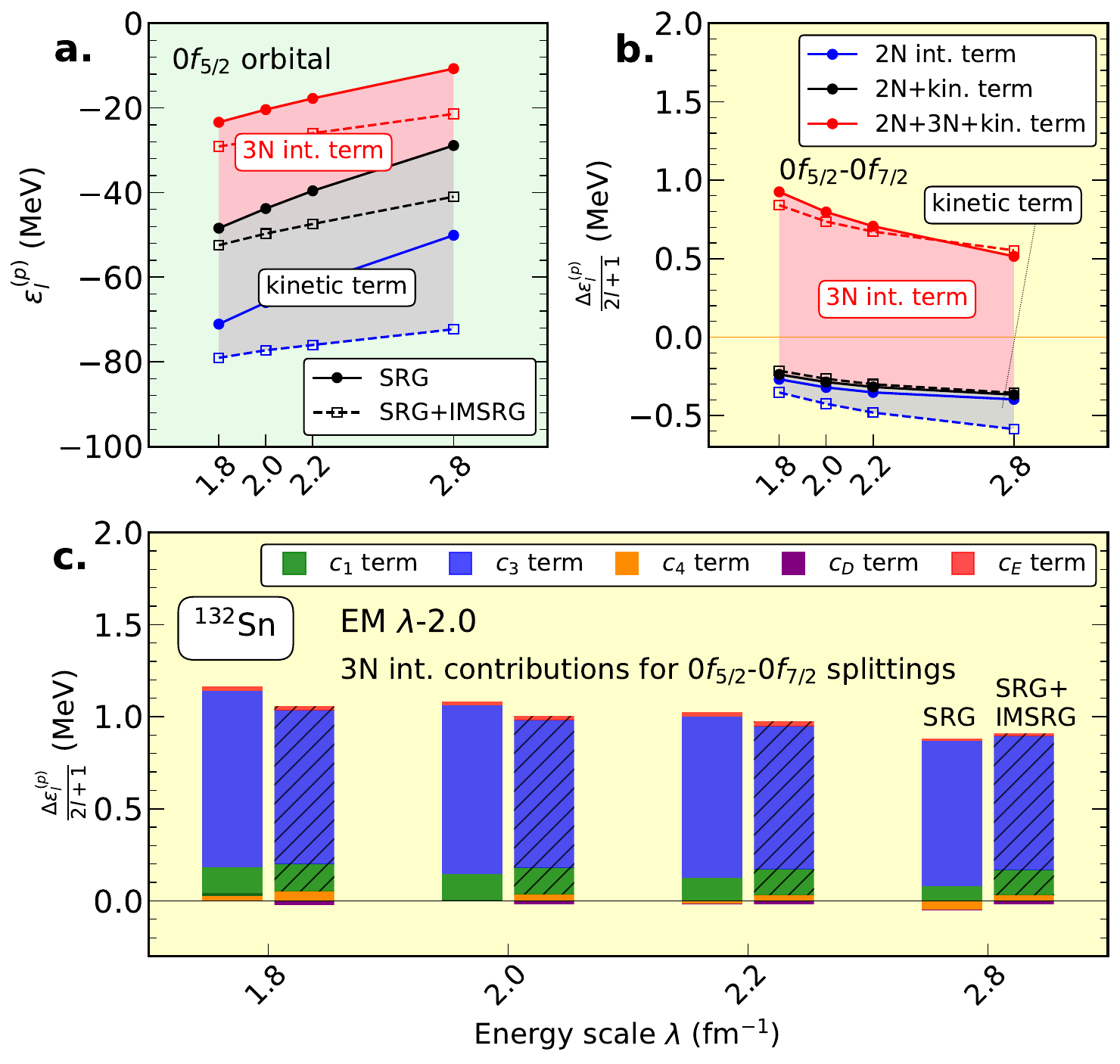} 
\caption{\textbf{Decomposition of the ESPEs and the SO splittings in \nuclide[132]{Sn} using the {\tt EM} family interaction as functions of the energy scale $\lambda$.}
\textbf{a}, The ESPE of the $0f_{5/2}$ state.
\textbf{b}, The energy splitting of the SO doublets $(0f_{5/2}, 0f_{7/2})$.
\textbf{c}, Contributions of different terms in the $3N$ interaction to the SO splitting shown in \textbf{b}.
}
\label{fig:EMdecomp}
\end{figure}
 
To test the robustness of the transition picture, we repeated the calculations using additional chiral interactions: the {\tt EM} interactions of ~\cite{Hebeler:2011}, as well as SRG-evolved versions of the $\Delta$-full {\tt N2LOGO(394)} interaction~\cite{Jiang:2020}. As shown in the SM, for the {\tt EM} family interactions, the results exhibit a similar trend to those obtained with the {\tt NN+3N(lnl)} interactions, albeit somewhat weaker. Since the {\tt N2LOGO(394)} interaction is inherently of low resolution, the SRG evolution does not significantly alter the shell structure as $\lambda$ decreases to 1.8 fm$^{-1}$, but the expected pattern is observed. 

\textit{\textbf{Connection between  nuclear forces and shell structure.}} 
Previous studies~\cite{Ando:1981,Pieper:1993} have demonstrated that $NN$ forces account for approximately half of the SO splitting in $p$-shell nuclei, while the other half arises from higher-body forces. Specifically, $3N$ forces play a crucial role in explaining the shell structure in calcium isotopes~\cite{Holt:2012,Hagen:2012nx,Hergert:2014,Soma:2014} and the oxygen dripline~\cite{Otsuka:2010prl,Hagen:2012oq,Cipollone:2013uq} in a low-resolution picture. The contribution of chiral $3N$ forces to the SO part of the EDF has also been studied in Refs.~\cite{Kaiser:2003,Kaiser:2010,Holt:2011}, showing that this contribution increases with the density of nuclear matter, and a recent study~\cite{Fukui:2024} suggests that the vector component of the $3N$ force is mainly responsible for enlarging the SO splitting of the $p$ orbitals. 

In our framework, we can identify how each component of $H_\lambda$ contributes to the ESPEs as a function of momentum resolution. As shown in Fig.\ref{fig:EMdecomp}(a), the dominant contribution arises from the $NN$ interaction, which provides the strong attractive potential binding nucleons together. Then, the kinetic term increases the ESPE of the $0f_{5/2}$ orbital by approximately 30 MeV, while the $3N$ term contributes additional repulsion which grows as $\lambda$ decreases. This trend is similar both with and without additional IMSRG evolution, whose effect lowers the energy of the $0f_{5/2}$ orbital. 

The impact of the components on the SO splittings is illustrated for the doublet $(0f_{5/2}, 0f_{7/2})$ in Fig.~\ref{fig:EMdecomp}(b), where we see that the $3N$ interaction is the dominant contribution. The negative contribution of the $NN$ interaction is the result of a cancellation among different two-body terms in Eq.(\ref{Eq:H}), and is therefore sensitive to the single-particle wave functions. We find that this contribution becomes positive if the three-body interaction is switched off. 

For the {\tt EM} interactions, the contribution of the $3N$ interaction to the SO splitting can be further decomposed into three parts: the $2\pi$ exchange, $1\pi$ exchange, and contact terms, which are determined by the LECs ($c_1$, $c_3$, $c_4$), $c_D$, and $c_E$, respectively. Details are provided in the SM. The results obtained from the SRG-evolved and SRG+IMSRG-evolved interactions are shown in Fig.\ref{fig:EMdecomp}(c). In both cases, the $2\pi$ exchange potential is the dominant contributor, while the $1\pi$ exchange and contact contributions are negligible. Within the $2\pi$ exchange potential, the component associated with the LEC $c_3$ has accounts for 80\% to 90\% of the total $3N$ contribution to the SO splitting. This result is consistent with the recent finding in Ref.~\cite{Fukui:2024}, where it was shown that the vector component of the $c_3$ term in an irreducible-tensor decomposition generates the one-body SO potential~\cite{Ando:1981}. 

\begin{figure}[tb] 
    \includegraphics[width=\columnwidth]{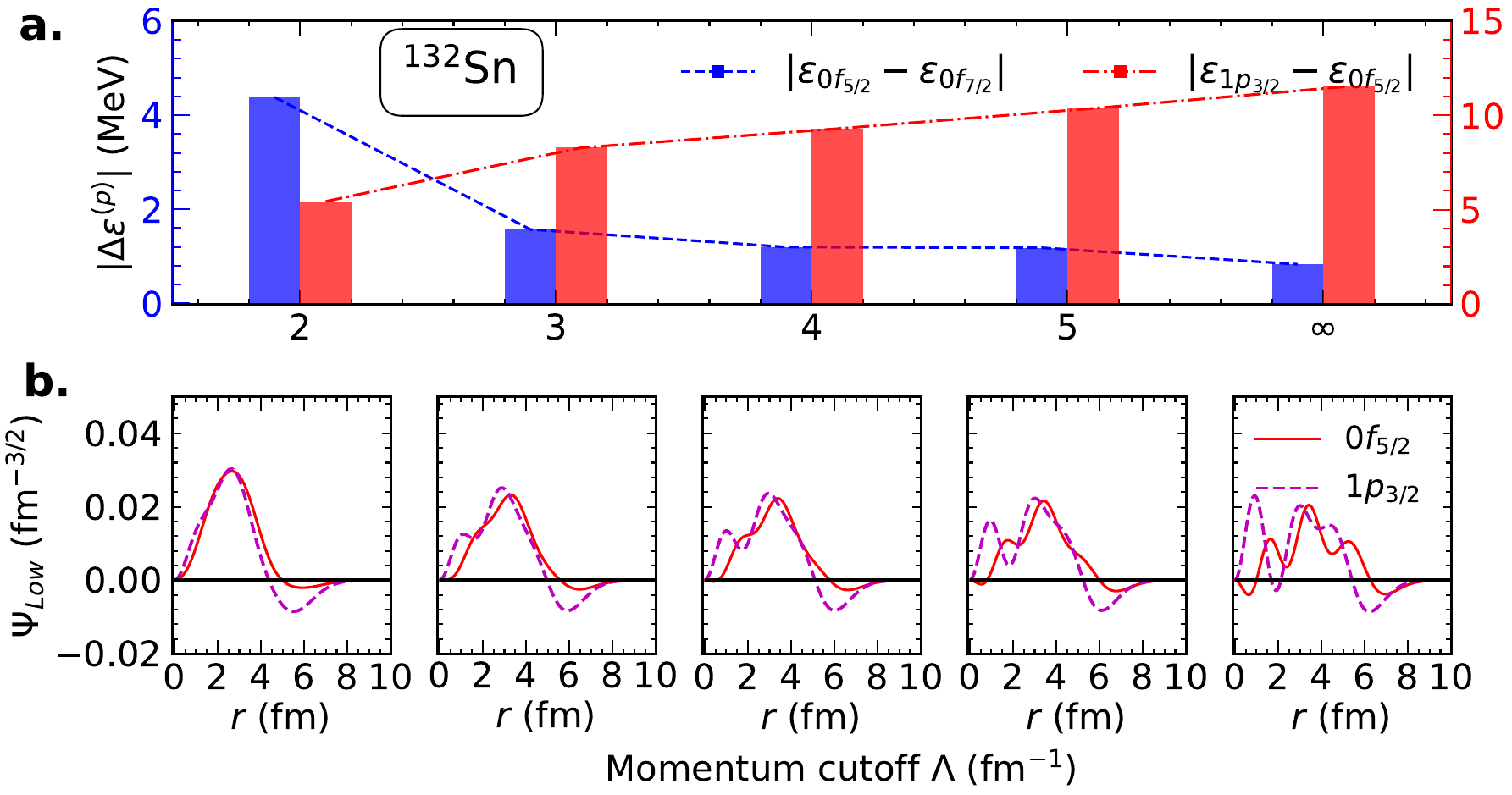}
    \caption{\textbf{Evolution of energy splittings $\Delta\varepsilon$ for SO and pseudospin doublets  of protons in \nuclide[132]{Sn} with the {\tt OBEP$\Lambda$} potentials.}  \textbf{a,} The energy splitting of the SO doublets ($0f_{7/2}, 0f_{5/2}$) and pseudospin  doublets ($1p_{3/2}, 0f_{5/2}$) as functions of the momentum cutoff $\Lambda$. 
    \textbf{b,} Evolution of the radial  wave functions of the  lower component of the Dirac spinors for the pseudospin doublets ($1p_{3/2}, 0f_{5/2}$).  The strong oscillations in the wave functions for the hardest potential ($\Lambda=\infty$) arise from the unbound nature of the states. }
    \label{fig:Sn132wf}
\end{figure}

Furthermore, as the resolution scale decreases from $\lambda=2.8$ fm$^{-1}$ to $\lambda=1.8$ fm$^{-1}$, the energy splitting between the SO partners $0f_{5/2}$ and $0f_{7/2}$ increases by a factor of 80\% (52\%) in the SRG (SRG+IMSRG) calculations. About 68\% (51\%) of the increase is contributed by the $3N$ force, whose contribution increases by 32\% (16\%) as the resolution scale is lowered. As illustrated in the SM, the enhancement of the SO splitting is responsible for the onset of the magic numbers $Z=28, 50$. Moreover, the pseudospin splittings primarily stem from the contributions of the kinetic energy term and the $NN$ interaction term, whose sizes decrease with $\lambda$.

 \textit{\textbf{Nuclear shell structure from relativistic studies.}} 
To test the universality of the transition picture, we extend the analysis to the shell structure generated by  relativistic OBEPs~\cite{Brockmann:1978,Shen:2019}. Evolving realistic nuclear forces to different momentum scales via SRG or IMSRG remains challenging in relativistic formulations; despite recent progress within the no-sea approximation~\cite{Zou:2024plb,Yang:2025cpl,Huang:2025plb}, a fully consistent implementation is still lacking and lies beyond the scope of this work. Instead, we constructed a family of relativistic OBEPs, regularized by a non-local cutoff $\Lambda$ ranging from $\infty$ to 2.0 fm$^{-1}$, with parameters fitted to $NN$ scattering data and the deuteron binding energy~\cite{Wang:2023zdc}. Using these nonlocal potentials, labeled as {\tt OBEP$\Lambda$}s, we performed RHF calculations~\cite{Wang:2025}. The key features of the ESPEs are highlighted for \nuclide[132]{Sn} in Fig.\ref{fig:Sn132wf}(a). As the momentum cutoff $\Lambda$ decreases, the energy splitting of the SO doublet ($0f_{7/2}, 0f_{5/2}$), which defines the magic number $Z=28$, increases, while the energy splitting of the pseudospin doublet ($1p_{3/2}, 0f_{5/2}$) decreases significantly. This again signals the evolution of nuclear shell structure from spin symmetry to pseudospin symmetry.  
Ginocchio first recognized that the ``pseudo'' orbital angular momentum corresponds to that of the lower components of the Dirac wave functions for pseudospin doublets~\cite{Ginocchio:1997}. Since then, the similarity of the lower radial wave functions of doublet partners has been regarded as a key signature of pseudospin symmetry~\cite{Meng:1999,Liang:2015}. As shown in Fig.~\ref{fig:Sn132wf}(b), the lower radial wave functions of the ($1p_{3/2}, 0f_{5/2}$) partners indeed become increasingly similar as $\Lambda$ decreases. An even more pronounced evolution trend is observed in $^{90}$Zr,  as shown in the SM.

 We note that RHF calculations with the {\tt OBEP$\Lambda$} potentials~\cite{Wang:2023zdc}, even without explicit $3N$ forces, yield a shell-structure evolution similar to that obtained with chiral $NN+3N$ forces. This suggests that, in the relativistic framework, some of the $3N$ effects may instead be obtained through virtual nucleon–antinucleon excitations~\cite{Brown:1987,Shen:2017}, which induce a density dependence during self-consistent RHF calculations through a dressed Dirac mass and energy in the lower component~\cite{Wang:2025}. Moreover, as shown in Ref.~\cite{Wang:2023zdc}, lowering the cutoff $\Lambda$ drives the meson–nucleon couplings $g_{\alpha \in \{\sigma, \pi, \omega, \delta, \eta, \rho\}}$ in {\tt OBEP$\Lambda$} towards those of effective interactions like PKA1~\cite{Long:2007_PKA}, which were optimized to reproduce empirical ground-state and shell-structure properties. In PKA1, the artificial shell gap at $Z=58$ disappears, and pseudospin symmetry is well preserved, reflecting the in-medium balance between the $\sigma$-meson attraction and $\omega$-meson repulsion, with important contributions from the $\rho$-meson tensor coupling~\cite{Geng:2019}. The present study thus provides new insight into the microscopic foundations of RMF models based on phenomenological effective interactions.
   
\textit{\textbf{Implications for shell structure in neutron-rich nuclei.}} The advent of radioactive ion beam facilities has enabled the production of exotic nuclei far from the $\beta$-stability line, sparking significant research interest in the evolution of nuclear shell structure toward the neutron dripline. To shed light on the shell structure in neutron-rich nuclei, we analyze the SO splittings in calcium isotopes — spanning the traditional magic numbers $N=20, 28$ and $50$ — using three different chiral Hamiltonians, as shown in the SM. The results clearly demonstrate that the SO splitting of proton states decreases monotonically with increasing neutron number from stable to unstable neutron-rich nuclei, signaling the gradual melting of traditional shell structures in atomic nuclei towards the neutron dripline. This finding also aligns well with the findings of self-consistent mean-field studies based on empirical EDFs~\cite{Dobaczewski:1994}.

 \textit{\textbf{Conclusions.}} The nuclear shell structure lies at the foundation of our understanding of atomic nuclei, governing their stability and shaping the elemental abundances observed in the Universe. A long-standing goal of nuclear theory is to derive this shell structure, including the emergence of magic numbers, directly from the underlying nuclear interactions. While previous studies have largely relied on phenomenological models  based on effective nuclear forces at low resolution scales, the connection between nuclear shell structure and realistic nuclear forces defined at high resolution has remained elusive. In this work, we have explored the evolution of shell structure as a function of momentum resolution, starting from realistic nuclear forces, including chiral two- and three-body interactions evolved via SRG and IMSRG, as well as relativistic OBEPs with varying momentum cutoffs.  We uncovered a novel transition from spin to pseudospin symmetry with decreasing resolution scale, accompanied by the natural emergence of the expected magic numbers. This transition is robust and universal, appearing consistently across different nuclei, interactions, and in both relativistic and nonrelativistic frameworks. By decomposing the contributions of various terms in the chiral Hamiltonians, we confirmed the essential role of the $3N$ force in enhancing spin-orbit splittings during the evolution. These results establish a direct and quantitative connection between high-resolution nuclear forces and the empirical effective nuclear potentials at a coarse-grained scale commonly used to describe nuclear shell structure.

\begin{acknowledgments}
\textit{\textbf{Acknowledgments.}}
We thank W. H. Long, B. N. Lu, J. Meng, T. Otsuka, S. H. Shen, and S. G. Zhou for fruitful discussions. We are especially grateful to B. C. He for his assistance with validating higher-order IMSRG effects.  This work is supported in part by the National Natural Science Foundation of China (Grant Nos. 12405143, 12375119, and 12141501) and the Guangdong Basic and Applied Basic Research Foundation (2023A1515010936). H.H. acknowledges the support of the U.S. Department of Energy, Office of Science, Office of Nuclear Physics,, under Award Numbers DE-SC0023516 and DE-SC0023175 (SciDAC-5 NUCLEI Collaboration). S.K.B. was supported in part by National Science Foundation (NSF) Grants PHY-2013047 and PHY-2310020.  
 
\end{acknowledgments}

%
 
%

\end{document}